\documentclass[prb,twocolumn,superscriptaddress,showpacs]{revtex4-1}

\usepackage{graphicx}
\usepackage{epstopdf}

\begin{document}

\title{Interstitial iron tuning of the spin fluctuations in Fe$_{1+x}$Te}

\author{Chris Stock}
\affiliation{NIST Center for Neutron Research, National Institute of Standards and Technology, 100 Bureau Dr., Gaithersburg, MD 20889}
\affiliation{Indiana University Cyclotron Facility, 2401 Milo B. Sampson Lane, Bloomington, IN 47404} 

\author{ Efrain E. Rodriguez}
\affiliation{NIST Center for Neutron Research, National Institute of Standards and Technology, 100 Bureau Dr., Gaithersburg, MD 20889}

\author{Mark A. Green}
\affiliation{NIST Center for Neutron Research, National Institute of Standards and Technology, 100 Bureau Dr., Gaithersburg, MD 20889}
\affiliation{Department of Materials Science and Engineering, University of Maryland, College Park, MD  20742}

\author{Peter  Zavalij}
\affiliation{Department of Chemistry, University of Maryland, College Park, MD  20742}

\author{Jose A. Rodriguez-Rivera}
\affiliation{NIST Center for Neutron Research, National Institute of Standards and Technology, 100 Bureau Dr., Gaithersburg, MD 20889}
\affiliation{Department of Materials Science and Engineering, University of Maryland, College Park, MD  20742}

\date{\today}

\begin{abstract}

Using neutron inelastic scattering, we investigate the low-energy spin fluctuations in Fe$_{1+x}$Te as a function of both temperature and interstitial iron concentration.  For Fe$_{1.057(7)}$Te the magnetic structure is defined by a commensurate wavevector of ($\frac{1}{2}$, 0, $\frac{1}{2}$).  The spin fluctuations are gapped with a sharp onset at 7 meV and are three dimensional in momentum transfer, becoming two dimensional at higher energy transfers.  On doping with interstitial iron, we find in Fe$_{1.141(5)}$Te the ordering wavevector is located at the (0.38, 0, $\frac{1}{2}$) position and the fluctuations are gapless with the intensity peaked at an energy transfer of 4 meV.  These results show that the spin fluctuations in the Fe$_{1+x}$Te system a can be tuned not only through selenium doping, but also with interstitial iron.  We also compare these results with superconducting concentrations and in particular the resonance mode in the Fe$_{1+x}$Te$_{1-y}$Se$_{y}$ system.

\end{abstract}


\maketitle

Magnetism is directly related to superconductivity in several heavy fermion and $d$-transition metal ion systems.~\cite{Bernhoeft98:81,Stock08:100}  Most notably, localized magnetism is believed to be directly coupled with high temperature superconductivity in the cuprates as evidenced through a series of detailed studies as a function of hole concentration where superconductivity is found to occur at a critical concentration of p$_{c}$=0.055, destroying long-ranged magnetic order.~\cite{Kastner98:70}    More recently, the discovery of superconductivity in the iron based materials have revealed a series of materials where superconductivity and magnetism coexist.~\cite{Johnston10:59}  Magnetism and superconductivity are strongly intertwined in these systems as illustrated through a series of neutron inelastic scattering studies which have presented a distinct change in the spin fluctuations on cooling through T$_{c}$.~\cite{Christianson08:456,Chistianson09:103,Lumsden09:102}  Arguably the simplest iron based superconductor is the layered Fe$_{1+x}$Te$_{1-y}$Se$_{y}$ system where superconductivity has been observed with a maximum T$_{c}$=14 K for $y \sim$ 0.5.~\cite{Sales09:79}

The magnetic structure of the parent non superconducting Fe$_{1+x}$Te has been investigated in powders using neutron diffraction and have reported the existence of a commensurate double stripe spin-density wave phase for low concentrations of $x$ with an ordering wave vector of {\bf{q}}$_{0}$=($\frac{1}{2}$, 0, $\frac{1}{2}$).~\cite{Bao09:102,Fruchart75:10}  For larger concentrations of iron, the magnetic phase becomes incommensurate along the $a^{*}$ direction and the structure is believed to be defined by a magnetic spiral.  The superconducting variants of Fe$_{1+x}$Te$_{1-y}$Se$_{y}$ have been investigated and neutron scattering has reported the static magnetic order observed in the parent material is replaced by short range magnetic correlations peaked near the {\bf{q}}$_{0}$=($\frac{1}{2}$,$\frac{1}{2}$,L) position.  This has led to the suggestion that the magnetic correlations shift from the ($\pi$,0) positions to the ($\pi$, $\pi$) points on becoming superconducting.  Most notably in the superconducting phase, a resonance peak at $\sim$ 7 meV has been observed in approximately half doped ($y \sim$0.5) systems near the {\bf{q}}$_{0}$=($\frac{1}{2}$,$\frac{1}{2}$,L) positions.~\cite{Liu10:9,Xu10:82,Babkevich10:22,Argyriou10:81}  The peak appears below the superconducting transition and is sharp in energy.  Unlike its counterpart in heavy fermion superconductors, the momentum dependence is very two-dimensional in nature, forming a rod along the $c^{*}$ direction. ~\cite{Qiu09:103} 

The microscopic nature of the magnetism and its relation to superconductivity in the FeAs-based high temperature superconductors is a matter of current debate and research.  While the reduced ordered moment on the iron site of $\sim$ 0.3-0.75 $\mu_{B}$ may indicate that itinerant effects are important, there are two key differences with the Fe$_{1+y}$Te system which may point towards strong localized magnetism in this system.   Firstly, the ordered moments are significant with values of 2.5 $\mu_{B}$ being reported for Fe$_{1.05}$Te.~\cite{Martinelli10:81}  Secondly, the ordering wave vector is not consistent with nesting wave vector measured by APRES.~\cite{Xia09:103}    Nevertheless, a strong coupling between the magnetism and electronics is implied by a series of optics studies which have observed a strong increase in the electronic lifetime below the magnetic ordering transition in Fe$_{1+x}$Te compounds.~\cite{Chen09:79}   

\begin{figure}
\includegraphics[width=9cm]{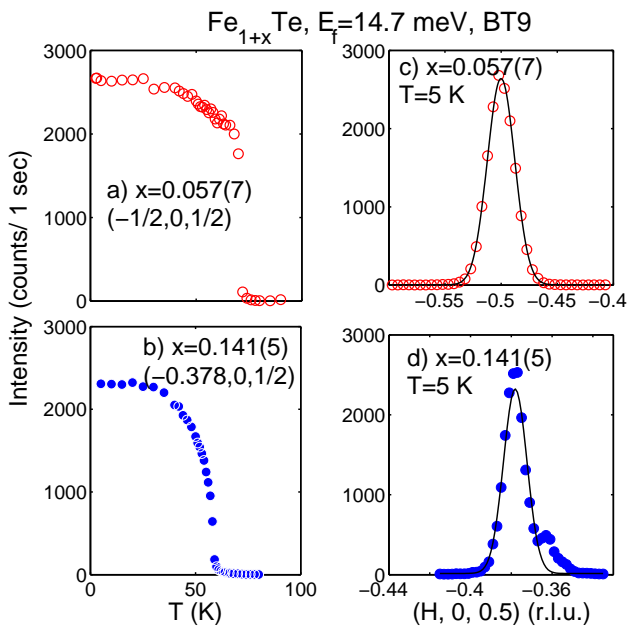}
\caption{\label{elastic} The temperature and momentum dependence of the elastic magnetic scattering measured in Fe$_{1.057(7)}$Te ($a$ and $c$) and Fe$_{1.141(5)}$Te($b$ and $d)$) taken on warming.  The data were taken with graphite filters and 80' collimators placed before and after the sample using BT9.  The solid lines represent fits to resolution limited gaussian line shapes.}
\end{figure}

Little attention has been placed on the role of excess Fe on the spin fluctuations and transport properties with much attention focussed on the effect of Se doping in the phase diagram.~\cite{Bendele10:82}  The importance of interstitial iron on the electronic properties has been highlighted by recent work which found for a fixed concentration of Se, the superconducting volume fraction could be independently tuned to zero with the introduction of interstitial iron.~\cite{Rodriguez11:xx}  Density functional calculations have further suggested that the excess iron plays a key role in the electronic properties with one electron carrier per excess Fe being added.~\cite{Zhang09:79}  Therefore, tuning interstitial iron may provide an alternate route for controlling the charge doping in FeTe layers.

To investigate the effect of interstitial iron on the low-energy spin fluctuations, we have performed neutron scattering studies on the parent Fe$_{1+x}$Te in the absence of superconductivity.   The 6 g single crystal samples were prepared by the Bridgeman technique. The best growth conditions for the Fe$_{1.057(7)}$Te crystal included melting the sample at 815$^{\circ}$ C for 12 hours, followed by a cooling rate of 6$^{\circ}$ C/hr.   The Fe$_{1.141(5)}$Te crystal was prepared with a similar heating time and cooling rate, but at a higher melt temperature of 850$^{\circ}$ C.  To prevent loss of iron content via reaction with the quartz ampoule, a pre-made powder sample of Fe$_{1.057(7)}$Te was mixed with excess iron powder to reach the Fe$_{1.141(5)}$Te stoichiometry.   Single crystal x-ray diffraction on crystals cleaved from the larger crystals was performed to characterize the amount of interstitial iron.

\begin{figure}
\includegraphics[width=8.3cm]{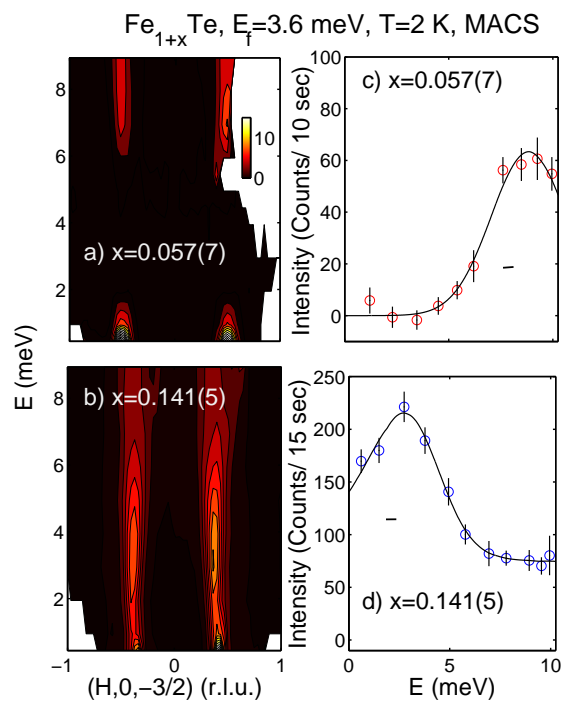}
\caption{\label{constQ} Constant-Q slices ($a$ and $b$) and constant-Q scans obtained using the MACS cold triple-axis spectrometer. $a$ and $b$ are integrated over L=[-1.525,-1.475].  Constant-Q cuts are displayed in panels $c$ and $d$.  Panel $c$ integrates over H=[-0.55,-0.45] and L=[-1.55,-1.45].  Panel $d$ integrates over H=[-0.4,-0.36] and L=[-1.55,-1.45].  The solid curves are guides to the eye.}
\end{figure}

The samples were then aligned in the (H0L) scattering plane and the spin fluctuations were mapped out using the MACS (Multi Axis Crystal Spectrometer) cold triple-axis spectrometer located at the NIST Center for Neutron Research (Gaithersburg, United States).  Instrument details and design concepts can be found elsewhere.~\cite{Rodriguez08:19}  Constant energy planes were scanned by fixing the final energy to E$_{f}$=3.6 meV using the 20 double bounce PG(002) analyzing crystals and detectors and varying the incident energy defined by a double focussed PG(002) monochromator.  Each detector channel is collimated using 90' soller slits before the analyzing crystal.    Full maps of the spin excitations in the (H0L) scattering plane as a function of energy transfer were then constructed by measuring a series of constant energy planes.  All of the data has been corrected for $\lambda/2$ contamination of the incident beam monitor and an empty cryostat background has been subtracted.~\cite{Shirane_book}

The elastic magnetic scattering, measured on BT9 thermal triple-axis spectrometer (Fig. \ref{elastic}), was used to characterize the magnetic properties of the two Fe$_{1+x}$Te samples.  The temperature dependence of the magnetic ordering is illustrated in panels $a)$ and $b)$ for Fe$_{1.057(7)}$Te and Fe$_{1.141(5)}$Te, respectively.  The ordering wave vector is illustrated through H scans shown in panels $c)$ and $d)$.  The iron poor sample (Fe$_{1.057(7)}$Te) displays a sharp first-order transition at 75 K on heating with commensurate ordering defined by Bragg peaks with the propagation wave vector of {\bf{q}}$_{0}$=($\frac{1}{2}$,0,$\frac{1}{2}$).  On doping with Fe, this transition (panel $b)$) becomes characterized by an incommensurate wavevector at H=0.38 (panel $d)$) and a lower transition temperature of 60 K.  We note that no evidence of a commensurate phase characterized by scattering near {\bf{q}}$_{0}$=($\frac{1}{2}$,0,$\frac{1}{2}$) was found in the Fe$_{1.141(5)}$Te sample and likewise no strong incommensurate scattering was observable in the commensurate Fe$_{1.057(7)}$Te sample.  Therefore, these two materials provide a clean representation of the magnetic properties in the commensurate and incommensurate phases of Fe$_{1+y}$Te.

The inelastic scattering measured on MACS is summarized at T=2 K in Fig. \ref{constQ}  through a series of constant-Q cuts taken by integrating over L=[-1.55,-1.45].  Panels $a)$ and $b)$ compare the excitations in the ordered state at low temperatures for Fe$_{1+y}$Te with $x$=0.057(7) and 0.141(5).    The excitations in the x=0.057(7) sample are gapped with a value of 7 meV, while those in the interstitial Fe rich x=0.141(5) concentration are gapless yet with the excitations peaked at around 4 meV as represented by the constant-Q cuts displayed in panels $c$ and $d$.    For both concentrations, the excitations are well correlated along the H direction indicating strong correlations in the $a-b$ plane.  The magnetic correlations form rods in energy with little dispersion over the energy range investigated in this experiment.   Studies using spallation neutrons have observed the scattering to extend up to energies greater than 250 meV indicating possibly an exceptionally large value for the coupling between Fe spins within the $ab$ plane and a strong role of next nearest neighbor interactions.~\cite{Lumsden10:6,Lipscombe11:106} 

Figure \ref{constQ} $c, d)$ illustrate constant-Q scans along the peak of correlated magnetic scattering.  The commensurate Fe$_{1.057(7)}$Te displays a distinct spin gap with a value of 7 meV and the incommensurate Fe$_{1.141(5)}$Te sample displays gapless excitations, although peaked at 4 meV.  These results illustrate that the characteristic energy can be tuned with interstitial iron and charge doping.

\begin{figure}
\includegraphics[width=9.6cm]{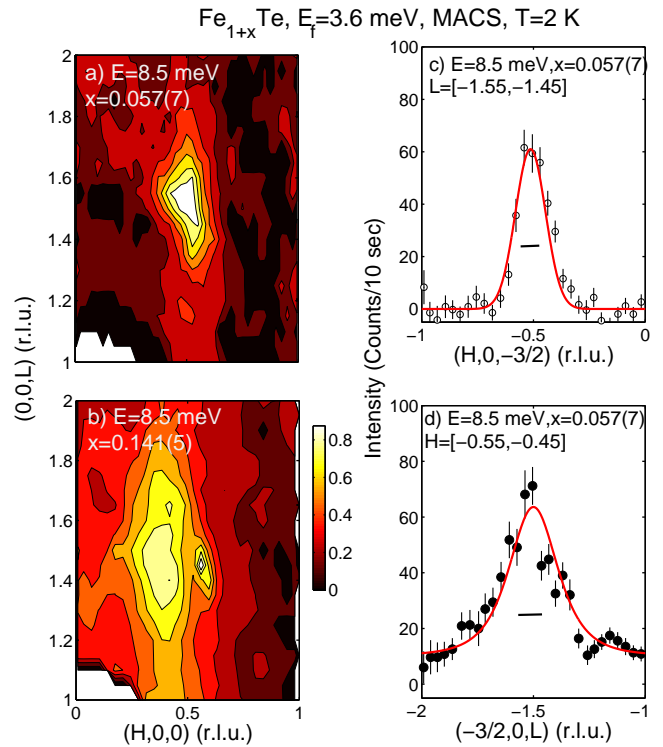}
\caption{\label{Q_cuts} Constant energy cuts at 8.5 meV in both the commensurate ordered ($x$=0.057(7)) and incommensurate ($x$=0.141(5)) ordered samples.  Panels $a)$ and $b)$ illustrate constant energy contours in the (H0L) scattering plane at 7.5 meV for the different interstitial iron concentrations.  Cuts through the correlated magnetic peak along the [100] and [001] directions are illustrated in panels $c)$ and $d)$ respectively for the commensurate $x$=0.057(7) sample.}
\end{figure}

The momentum dependence of the low-energy fluctuations in commensurate Fe$_{1.057(7)}$Te (panel $a)$) and incommensurate Fe$_{1.141(5)}$Te (panel $b)$) are summarized in Fig. \ref{Q_cuts} at energy transfers of 8.5 meV.   Constant energy contours are presented in panels $a)$ and $b)$ where it can be seen that the excitations are characterized by a well defined peak in the H and L directions.   The three dimensional character is further demonstrated by lines scans (taken from the $x$=0.057(7) sample) in panels $c)$ and $d)$ with the peaks centered near the commensurate {\bf{q}}$_{0}$=($\frac{1}{2}$,0,$\frac{1}{2}$) position.  However, while the correlations are well defined in both directions, the magnetic scattering is nearly resolution limited along the H direction but is significantly broader along L.  A comparison between panels $a)$ and $b)$ is further suggestive that the correlations along the c-axis weaken in the incommensurate $x$=0.141(5) phase.  The c-axis correlations in superconducting Fe$_{1+x}$Te$_{0.5}$Se$_{0.5}$ have been found to further weaken with the scattering in those samples to be described by nearly a rod of scattering along L.  

\begin{figure}
\includegraphics[width=9.4cm]{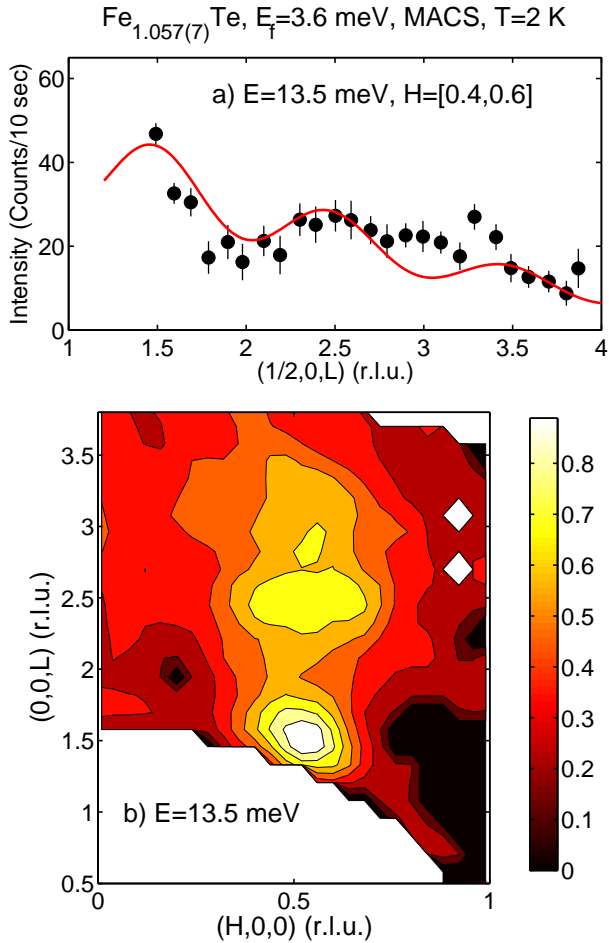}
\caption{\label{L_highE}  Constant energy cuts of the correlated magnetic intensity at 13.5 meV.  The lower panel $b)$ illustrates a contour plot of the magnetic fluctuations along the (0.5,0.5,L) direction and panel $a)$ plots a cut through the data along L.  The solid curve is a fit to nearest neighbor correlations demonstrating the two-dimensionality of the fluctuations as discussed in the text. }
\end{figure}

The scattering at higher energy transfers is illustrated in Fig. \ref{L_highE} which illustrates a constant energy contour in panel $b)$ at 13.5 meV for commensurate ordered Fe$_{1.057(7)}$Te.  The scattering is significantly broaden along both H and L and panel $a)$ demonstrates the magnetic correlations along L have broadened significantly, in particular in comparison to the data taken at 8.5 meV.  The solid line is a fit to the function $I(L)=A f^{2}(Q)(1+2\alpha\cos(2\pi L + \gamma))$ with $A$ and amplitude, $f(Q)$ the magnetic form factor for Fe$^{2+}$, $\alpha$ the correlation function between nearest neighbor layers, and $\gamma$ the phase of the spins in one layer with respect to the next taken to be $\pi$ as expected for antiferromagnetic correlations.  The function represents spins correlated with nearest neighbor layers only and further correlations can be accounted for by extending the Fourier series.   The fit demonstrates that an adequate description of the data can be obtained by considering nearest neighbor correlations only and that at high-energy transfers, the magnetic correlations become strongly two-dimensional.  This result is expected as the layers in Fe$_{1+x}$Te are bonded along the c-axis by weak Van der Waals forces mostly.

\begin{figure}
\includegraphics[width=8.7cm]{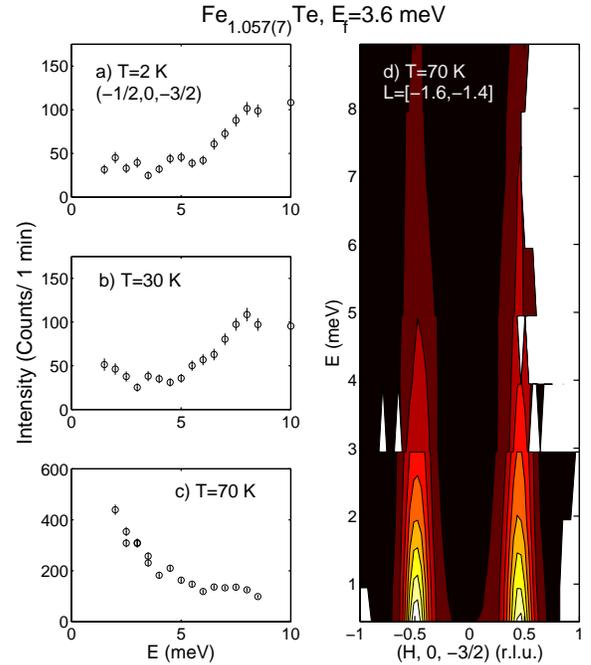}
\caption{\label{temp}  The temperature dependence of the magnetic excitations in Fe$_{1.057(7)}$Te.  The constant-Q scans in $a-c$ where taken with a single detector channel on MACS at T=2, 30, and 70 K.  $d$ illustrates a constant-Q slice taken at 70 K  illustrating the gapless excitations.}
\end{figure}

The temperature dependence of the spin fluctuations in Fe$_{1.057(7)}$Te  are illustrated in Fig.\ref{temp}  which plots a series of constant-Q scans taken with a single MACS detector channel ($a-c$) and a constant-Q slice taken at 70 K ($d$) integrating over L=[-1.6,-1.4].  The constant-Q scans illustrate a gradual filling in of the low-energy spin-fluctuations with increased temperature up to T$_{N}$, where the fluctuations are gapless but well correlated in H.  We note that we do not observe the gap to soften, but rather magnetic fluctuations at low-energies to build up and fill in the gap with increasing energy transfer.  At temperatures near the T$_{N} \sim$ 70 K, the gap is nearly completely filled in and the static magnetic order is destroyed (Fig. \ref{elastic}).

The main result from this work is the characterization of the low-energy spin fluctuations in Fe$_{1+x}$Te as a function of interstitial iron.  The comparison displayed in Fig. \ref{constQ} illustrate the dramatic effects that interstitial Fe, and the subsequent charge doping, have on the low-energy spin dynamics.  Such effects are likely present in superconducting concentrations of Fe$_{1+x}$Te$_{1-y}$Se$_{y}$ as interstitial iron clearly results in a filling in the spin excitations at low energies.   Such low-energy magnetic fluctuations are possibly destructive to superconductivity, and the interstitial iron concentration needs to be characterized in superconducting materials.  

It is interesting to compare our results with the magnetic fluctuations in superconducting Fe$_{1+x}$Te$_{1-y}$Se$_{y}$.  In particular, the commensurate ordered Fe$_{1.057(7)}$Te sample displays gapped three dimensional fluctuations.  The gap value has an identical energy scale to that observed in superconducting Fe$_{1+x}$Te$_{1-y}$Se$_{y}$ (optimal T$_{c}$=14 K) samples and also in the analogous BaFe$_{1.85}$Co$_{0.15}$As$_{2}$ (T$_{c}$=22 K) superconductor (denoted as the 122 system).~\cite{Qiu09:103,Lumsden09:102,Inosov10:6}   While the gap value has not been directly observed to scale with the superconducting temperature in Fe$_{1+x}$Te, the spin gap in the 122 system does appear to strongly scale with T$_{c}$ and therefore likely the superconducting gap.  While the spin gap in all iron based systems has been interpreted as a resonance peak in analogy to the cuprates, this interpretation may need reconsideration in the $``$11" Fe$_{1+x}$Te system as interstitial iron and effects due to localized magnetism are playing a strong role in the dynamics. 

The fact that a similar excitation spectrum exists in a non superconductint parent material implies that the gap value may not be directly related with the superconducting gap nor condensation energy as suggested based on the scaling of the resonance energy with T$_{c}$ in some cuprates.~\cite{Dai01:63} The presence of a $``$resonance" peak (which appears as gapped spin fluctuations) in the non superconducting parent compound is consistent with several theories proposed to explain the resonance peak in the cuprates and heavy fermion compounds.  The spin-fermion model describes the resonance peak in terms of a spin excitation which is broadened in energy in the normal state owing to the interaction between spin fluctuations and electronic channels.  These decay channels are then gapped in the superconducting phase, removing the dampening and resulting in sharp excitations.~\cite{Morr98:81}   Another theory describes the resonance in terms of a spin exciton and such a theory could explain the relatively small fractional amount of spectral weight which resides in the resonance peak.~\cite{Eremin05:94}   Our results may support the idea that the resonance peak in superconductors originates from a spin fluctuation, which is heavily damped in the normal state owing to electronic interactions.  In the parent compound, such dampening maybe absent owing to the fact that the ordering wavevector $(\pi,0)$ does not match any nesting wave vector which is the case for fluctuations near $(\pi, \pi)$ observed in superconducting systems.  Further investigations would be necessary to confirm this conjecture.

Despite the similarity between our measurements on non superconducting Fe$_{1.057(7)}$Te and superconducting Fe$_{1+x}$Te$_{1-y}$Se$_{y}$, there are several noteworthy differences.  Firstly, the gapped excitations in Fe$_{1.057(7)}$Te appear near {\bf{q}}$_{0}$=($\frac{1}{2}$,0,$\frac{1}{2}$) while superconducting concentrations display correlations near {\bf{q}}$_{0}$=($\frac{1}{2}$,$\frac{1}{2}$,L).  The later position in momentum has been suggested to correspond to a nesting wavevector of the Fermi surface which might imply stronger coupling between spin and electronic degrees of freedom for Se doped and superconducting materials.~\cite{Xia09:103,Lee10:81}  Secondly, the spin excitations in Fe$_{1.057(7)}$Te display strong three dimensional correlations at low-energies with two-dimensional fluctuations only present at energies nearly twice the value of the spin-gap.   This implies that the reduced dimensionality maybe a key factor for superconductivity in the Fe based materials and a such a scenario has been theoretically suggested for heavy fermion CeXIn$_{5}$ systems.~\cite{Monthoux02:66}

It is interesting to compare the case here with that observed in the CaFe$_{2}$As$_{2}$ and BaFe$_{2}$As$_{2}$ systems which becomes superconducting on doping with both K for the alkaline metal site and Co for the Fe site.  In both CaFe$_{2}$As$_{2}$ and BaFe$_{2}$As$_{2}$, the excitations display a gap of $\sim$ 10 meV and the spin-waves are strongly dispersive along H and L up to energy transfers greater than 50 meV.~\cite{Diallo09:102,Matan09:79}  This clearly contrasts with Fe$_{1+x}$Te where the fluctuations are two-dimensional in nature at much lower energy transfers.  The spin-wave gap, however, is very similar in Fe$_{1+x}$Te and (Ba,Ca)Fe$_{2}$As$_{2}$.  In superconducting phases of(Ba,Ca)Fe$_{2}$As$_{2}$ doped with K, Co, and Ni, the energy scale defined by the resonance energy, is tuned through charge doping.  In the situation with Fe$_{1+x}$Te, it appears that the low-energy scale is also tuned with charge doping from interstitial iron.     

In summary, we have presented a study of the low-energy spin fluctuations in the commensurate and incommensurate phases of the parent Fe$_{1+x}$Te material.  We have observed a distinct spin-gap in the commensurate material which occurs at the same energy scale observed for the magnetic resonance in superconducting Fe$_{1+x}$Se$_{y}$Te$_{1-y}$ concentrations, although at a different wave vector.  We also observe that this gap is filled in and replaced by gapless excitations on doping with interstitial iron and increasing temperature.  These results illustrate the importance of interstitial iron in understanding the role of magnetism in Fe based superconductors and that the connection of the resonance with the superconducting gap needs to be revaluated in these materials.

We are grateful to N. C. Maliszewskyj for expert technical support during these experiments.  This work utilized facilities supported in part by the National Science Foundation under Agreement No. DMR-0944772.

\end{document}